\documentclass[a4paper]{jpconf}
\usepackage{graphicx}
\begin{document}
\title{Ultra-peripheral nuclear collisions}

\author{Carlos A. Bertulani}

\address{Physics and Astronomy Department,  Texas A\&M University-Commerce, Commerce, Texas 75025, USA}

\ead{carlos.bertulani@tamuc.edu}

\begin{abstract}
This article presents a very brief review of the physics of Ultra-Peripheral Collisions (UPC) at the Large Hadron Collider (LHC) and other nuclear facilities. I discuss several processes of interest such as electron-position pair production, the anti-hydrogen atom, giant resonances, exotic meson production and parton distribution functions.
\end{abstract}

\section{Introduction}
In 1985, my PhD adviser (Gerhard Baur, J\"ulich, Germany) and I came up with a paper on relativistic Coulomb excitation \cite{BeBa85}. We showed that when the electromagnetic fields of one nucleus induces inelastic transitions in the other, the excitation process involves the same transition matrix elements as those induced by real photons.  These collisions are known as {\it Ultra-Peripheral Collisions }(UPC). In first-order perturbation theory, a factorization can be carried out leading to a sum over multipoles  \cite{BeBa85}, 
\begin{equation}
\sigma = \sum_{E/M,L}\int d\omega {n_{E/M,L}(\omega) \over \omega} \sigma_\gamma^{(E/M,L)} (\omega), \label{epn2}
\end{equation}
where $\sigma_\gamma^{(E/M,L)} (\omega)$ are the cross sections induced by real photons with energy $\omega$. The electric (E) and magnetic (M) multipolarities contain the photon angular momentum $L$ component. The functions $n_{E/M,  L} $ depend upon the projectile bombarding energy $E_{beam}$ and the excitation energy $\omega$. They are known as {\it equivalent (virtual) photon numbers} (EPN)  \cite{BeBa85}. At projectile energies below a few GeV/nucleon, the EPNs are strongly dependent on the multipolarity, e.g., $n_{E2} > n_{E1} > n_{M1}$ whereas at much large energies they are approximately equal, $n_{E2} \sim n_{E1} \sim n_{M1}$, except for the very low excitation energies $\omega\ll \gamma/b$ \cite{BeBa85}. Here, $\gamma = (1-v^2)^{-1/2}$ is the Lorentz factor, $v$ the relative velocity between the nuclei and $b$ the impact parameter in the collision.

Enrico Fermi formulated a similar idea already in 1924 \cite{Fer24,Fer25}  for the ionization of atoms by  fast $\alpha$-particles. His analytical result looks like Eq. (\ref{epn2}), but is only valid for the E1 multipolarity. Fermi's method is known as {\it Weisz\"acker-Williams (WW) method}, who introduced relativistic corrections in Fermi's formalism in 1934 \cite{Wei34,Wil34}. Intriguingly, Fermi published the same paper in two different journals in different languages (German in {\it Z. Phys.} and Italian in {\it N. Cimento}). Nobody really knows why, but one guesses that the reason is that Fermi was a young physicist looking for a faculty position in Italy and the version of his paper in {\it Nuovo Cimento} was intended to advertise his relatively unknown name in Italy at the time\footnote{Another possible explanation is that Nuovo Cimento, founded in 1923, in the beginning was not a real journal but collected reports from members of the Italian Physical Society (SIF) - I am indebted to Angela Bonaccorso for this clarification.}.   

At the Large Hadron Collider (LHC) at CERN, Switzerland, the $\gamma\gamma$ collisions induced by the electromagnetic (EM) fields of the colliding ions occur at center of mass energies an order of magnitude higher than were available at now defunct $e^+e^-$ colliders, and $\gamma$-nucleus collisions reach 30 times the energies available at fixed target accelerators. The Lorentz boost factor $\gamma$ in the laboratory frame is about 7000 for p-p, 3000 for Pb-Pb collisions. Because of the large charge of Pb ions ($Z=82$) and the short-interaction time, $\Delta t \simeq (10^{-20}/\gamma$) seconds, the EM  are stronger ($\propto Z^{2}$) than the Schwinger critical field \cite{Sch49,Bau08} $E_{Sch} = m^{2}/\hbar e = 1.3 \times10^{16}$ V/cm. Light
particles such as $e^{+}e^{-}$-pairs are produced copiously by such fields \cite{BeB88}.  

This brief review provides an incomplete account of the development of UPCs since my PhD thesis was published in 1988 \cite{BeB88}. It is rewarding to see that, despite the skepticism reigning when my thesis was published,  all the predictions we made in numerous works in the 1980s and 1990s  have motivated a flurry of experimental results and theoretical developments.  The physics of UPCs has become a subject of intensive studies worldwide and helps to consolidate numerous phenomena in QED and QCD \cite{KGS97,BHT02,BKN05,BHT07,Ba08,Ba08b,Nyst08}.

\section{Misconceptions}  Because of its popularity in dealing with high energy processes, it is often assumed that the EPN formulated for UPCs is only valid in relativistic heavy ion collisions. That is wrong. Eq. (\ref{epn2}) always holds if first-order perturbation theory is accurately enough to describe a low energy UPC process. This is a consequence of the assumption that during the excitation process $\nabla \times {\bf B} =0, \ \nabla \cdot {\bf E} =0$, where ${\bf E}$ and ${\bf M}$ are the EM fields generated by the source nucleus. 

Another common misconception is that one cannot separate pure EM processes from those induced by the strong interaction. Numerous ways have been devised to separate the two processes. One often utilizes angular distributions, excitation energies, laboratory energy and target dependence of the cross sections, etc. In particular, measurements using light (e.g., carbon) concurrently with heavy (e.g., lead) nuclei helps to separate the contribution of the two interactions.  
\begin{figure}[h]
\begin{minipage}{15pc}
\includegraphics[width=15pc]{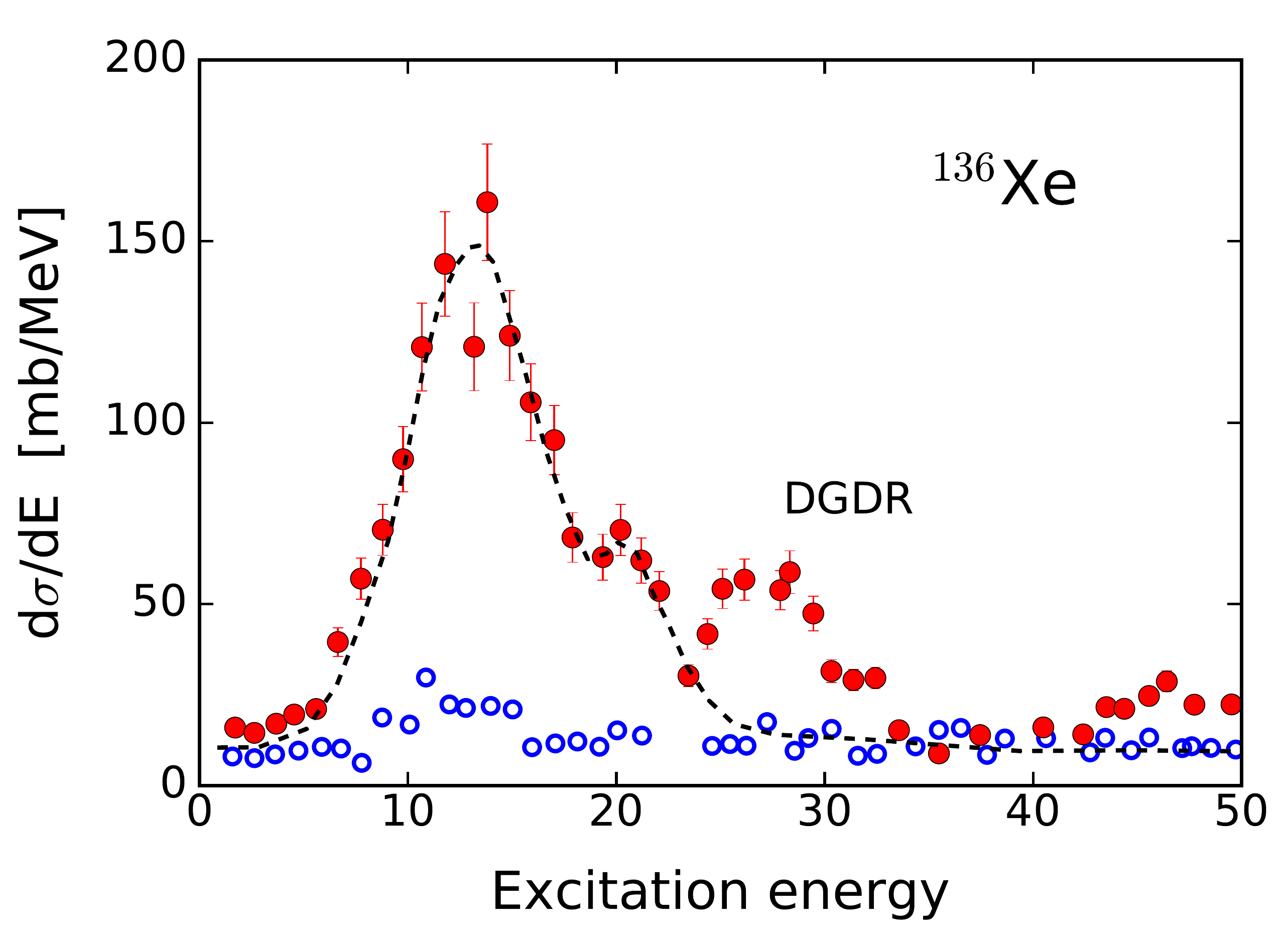}
\caption{Excitation spectrum of giant resonances in $^{136}$Xe in UPCs with a large-Z target. The DGDR was observed as a clear bump in the spectrum \cite{Sch93}. \label{dgdr}}
\end{minipage}\hspace{1pc}%
\begin{minipage}{11pc}
\includegraphics[width=11pc]{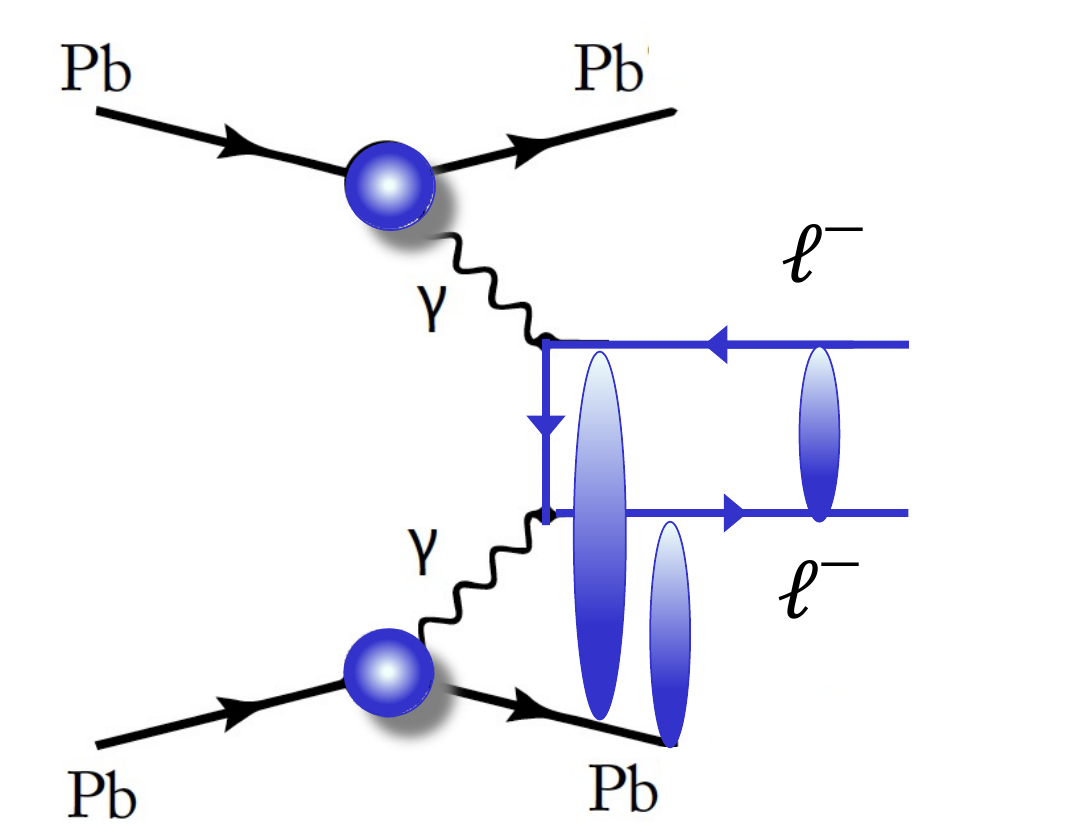}
\caption{\label{llpair}Lepton-antilepton pair production in UPCs. The blobs represent possible higher-order processes.}
\end{minipage}\hspace{1pc}%
\begin{minipage}{9pc}
\includegraphics[width=9pc]{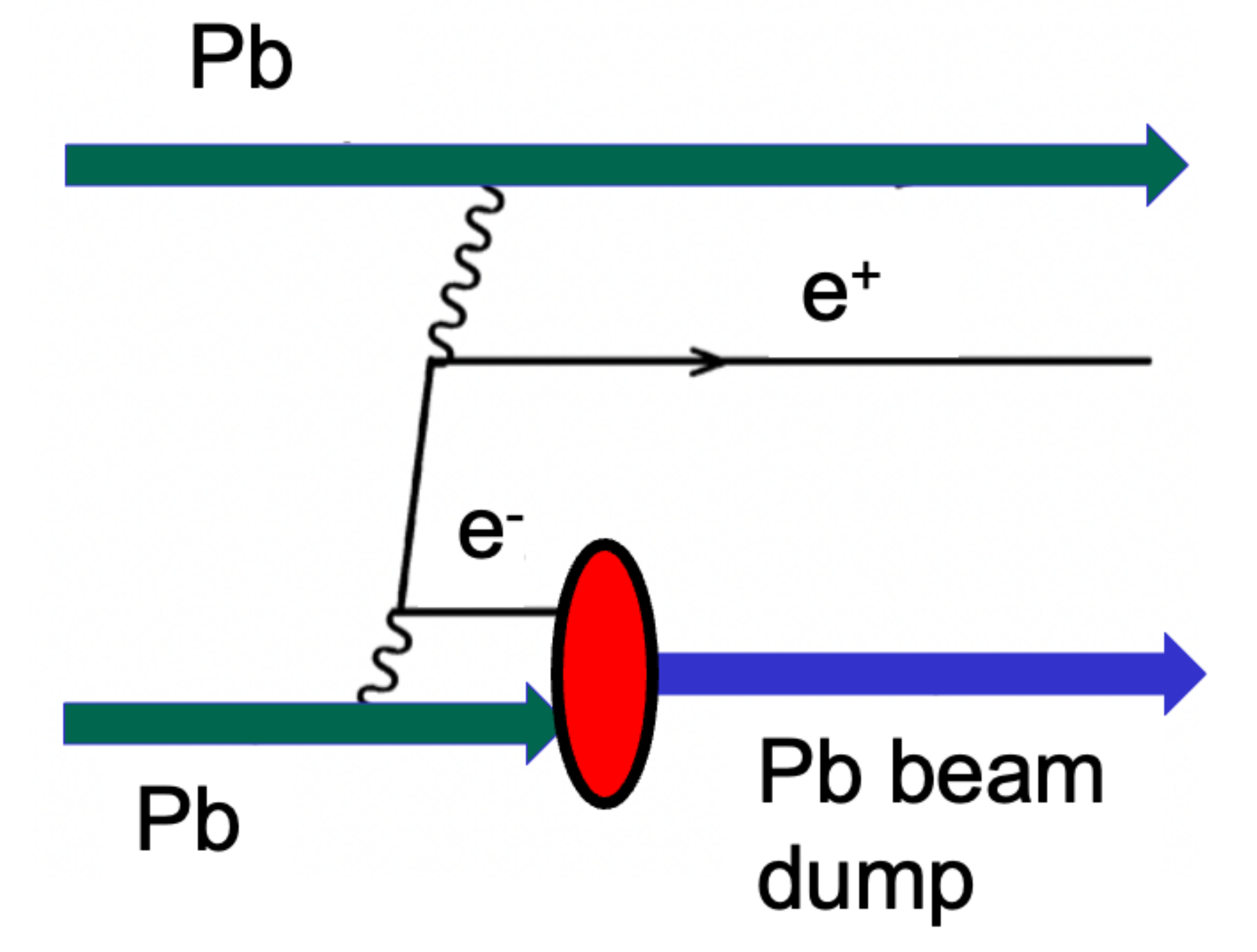}
\caption{\label{llpairc}Pair production with electron capture leading to possible beam losses \cite{BeB94}.}
\end{minipage}%
\end{figure}

\section{Giant Resonances} One of our first applications of UPCs was in the study of giant resonances in nuclei \cite{BeB86}. The excitation is most often followed by neutron emission and the cross sections are several barns for heavy nuclei even at moderate energies of $\sim 1$ GeV/nucleon such as those available at the GSI laboratory in Germany \cite{ABS95,ASBK96}. At these bombarding energies giant resonances can also be excited with a large probability by means of nuclear interactions. But, it was soon realized \cite{BeB88} that for large-Z nuclei the cross sections are much smaller than those induced by the EM interaction. Nowadays, Coulomb excitation and decay of giant resonances is a useful experimental probe, including nuclear fission studies \cite{Aum97,BKL20}. Due to its large cross section, the excitation and decay of giant resonances has been proposed as a luminosity monitor in heavy-ion colliders \cite{BCW98}.

 \section{Double giant resonances} The large probabilities for excitation of giant resonances led to our proposal in 1986  \cite{BeBa86,BeBau86} that multiple giant resonances could also be excited via higher-order processes, or multiple photon exchange.  When higher-order effects are of relevance, coupled-channels equations are one possible way to treat the process non-perturbatively. One can use the set of coupled-channel equations ($\hbar = c = 1$),
\begin{equation}
{i}\frac{d}{d z}{\cal A}_n({\bf b},z) = \sum_{m}\left\langle
\Phi_{n}|{\cal O}_{E/M,L}({\bf b},z)|\Phi_{m}\right\rangle
{\cal A}_{m}({\bf b},z)\ e^ {i \omega_{nm}  z }, \label{cceq4}
\end{equation}
where ${\cal A}_n({\bf b},z)$ is the excitation amplitude for a collision at the impact parameter ${\bf b}$, $z$ is the longitudinal coordinate, $v$ is the projectile velocity, ${\cal O}_{E/M,L}$ is the appropriate electromagnetic operator, and  $n,m$ denote nuclear states. Diffraction effects induced by nuclear interactions can also be included with help of Glauber methods  for nuclear absorption.  We predicted large cross sections for the excitation of double, triple, and multi-phonon resonances in nuclei and proposed experimental measurements \cite{BeBa86,BeBau86}, accomplished in two pioneer experiments at the GSI laboratory in Germany in 1993  \cite{Sch93,Rit93}. One experiment used gamma-gamma coincidences for the decay of the double giant dipole resonance (DGDR)  \cite{Rit93} and another looked at its particle (neutron emission) decay \cite{Sch93}. 

In the 1980s, nuclear structure experts told us that it would be impossible to identify the DGDR in UPC experiments because it would be immersed in a large background of many-particle-many-hole excitations. History proved once again that there is nothing better than experimenting and observing. Figure \ref{dgdr}  shows the  results of the experiment reported in  Ref. \cite{Sch93} for the DGDR excitation in $^{135}$Xe projectiles in UPCs obtained at the GSI in 1993. Among the many reasons why the DGDR is worth studying is because its strength and width are not well understood and can  be used to constrain nuclear models for the response to EM operators.  Extensive discussions of these models have been published in Refs.  \cite{Em94,ABE98,BP99}. 

\section{Electron-positron pair production} Electron-positron pairs (Fig. \ref{llpair}) are copiously produced in UPCs. The first papers for the production of $e^+e^-$ pairs in UPCs date back to  the 1930s. Bethe, Racah, Bhabha, Tomonaga, Nishina, Furry, and many others, developed methods to calculate pair production using the newly born Dirac equation. Dirac predicted the existence of the positron when he formulated his famous equation for the description of the dynamics of electrons. The positron would manifest as a hole in the ``vaccuum sea'' of electrons.  At the time the only way to look for this ``hole'' (positron) would be  in UPCs with cosmic rays with large kinetic energies $E$.
Assuming that the energy of the produced pairs are much larger than the electron rest mass, $m_e$, all the theoretical predictions yield a production cross section to leading order equal to $Z_1^2Z_2 \alpha^2 \ln^3 \gamma$, where the Lorentz factor is $\gamma \simeq E/m_e$. In 1986, we revisited these calculations using  Quantum Electrodynamics (QED), a practical formalism outside the reach of theorists in the 1930s. We have formulated a theory including final state interactions (the blobs in Fig. \ref{llpair}) using Bethe-Maximon distorted waves \cite{BeB88}. Because the cross sections are very large, of the order of {\it kilobarns} for a typical collider energy such as the LHC, we have shown that the role of higher-order corrections was worth studying \cite{BeB88,BeB89}. 

\begin{figure}[h]
\begin{minipage}{17pc}
\includegraphics[width=17pc]{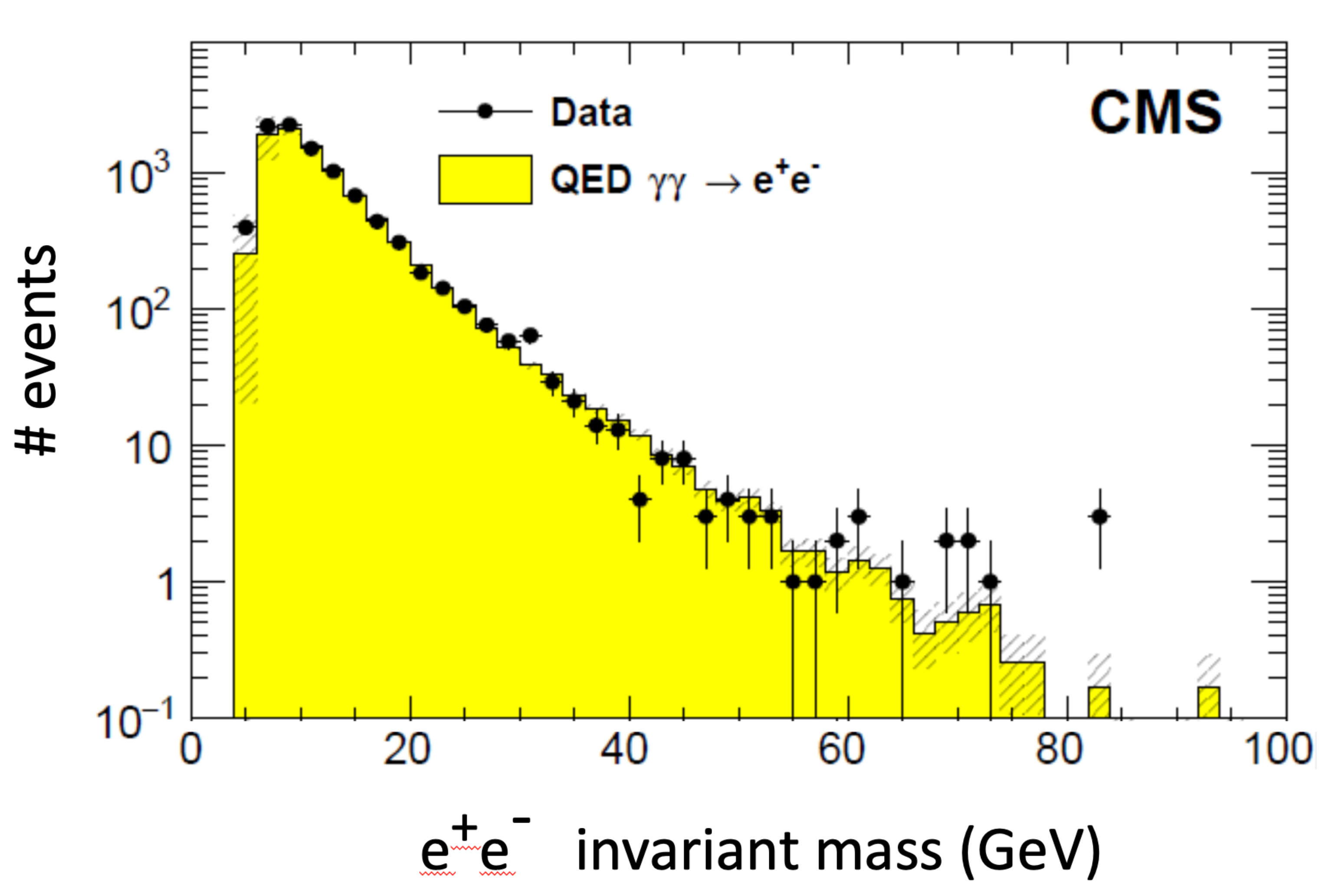}
\caption{\label{cms9}UPC production of $e^+e^-$ pairs as observed in the CMS detector at the LHC (adapted from Ref. \cite{Ad18}).}
\end{minipage} \hspace{1pc}
\begin{minipage}{19pc}
\includegraphics[width=19pc]{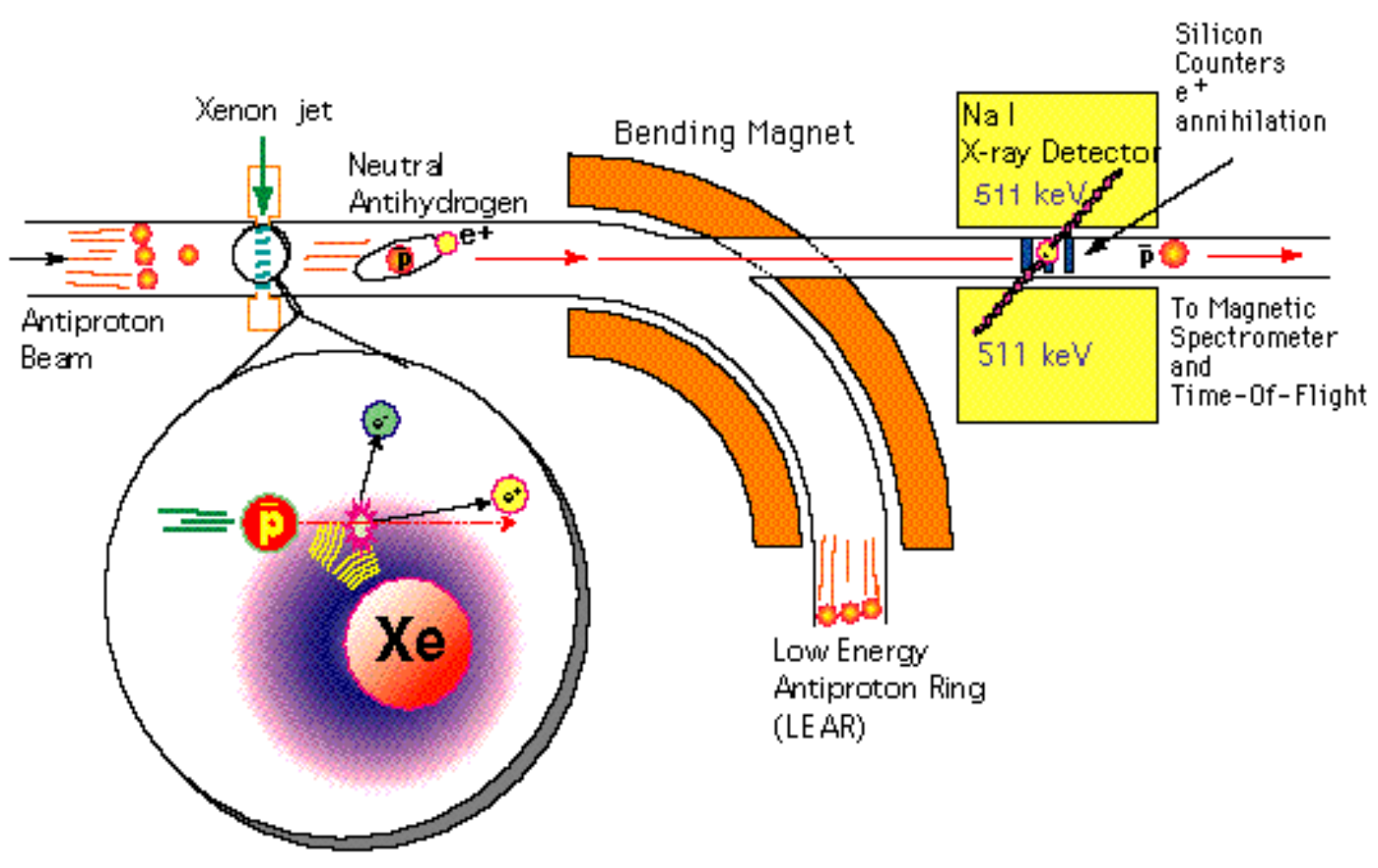}
\caption{\label{setup}Schematic view of the setup used to identify anti-atoms at the LEAR/CERN in 1996 \cite{Bau96}.}
\end{minipage} %
\end{figure}

It is delightful to see that current experiments at the LHC are measuring the very fundamental pair-production cross sections first calculated in the 1930s. They are now almost fully understood theoretically \cite{Ad18} (see Fig. \ref{cms9}). During the first experiments, the predictions based on QED were confirmed (Fig. \ref{cms9}, adapted from Ref. \cite{Ad18})). The production of other particle-antiparticle pairs such as  lepton-antilepton pairs (Figure \ref{llpair}), $\gamma \gamma\rightarrow \mu^{+}\mu^{-}$, $\gamma
\gamma\rightarrow \pi^{+}\pi^{-}$, $\gamma \gamma\rightarrow W^{+}W^{-}$, etc,  with such a mechanism is not negligible, as first found in Ref.  \cite{BeBa87,BeB88,Bau88,BeB94}.

\section{Anti-atoms} Arguably, the most beautiful application of UPCs is the production of anti-hydrogen via pair creation with the capture of the corresponding lepton by one of the colliding nuclei (Fig. \ref{llpairc}) \cite{BeB88}. The application of this method to create anti-hydrogen atoms at the LHC was proposed in Ref. \cite{MB94} and first observed by an ingenious experiment performed at CERN and published in 1996 \cite{Bau96}. At the Low Energy Antiproton Ring (LEAR) at CERN one used antiprotons colliding with protons and looked for the positron being captured in an atomic orbit around the antiproton.  It was the first time anti-atoms were  produced in a terrestrial laboratory with 11 anti-hydrogen atoms observed (see Fig. \ref{setup}). The exciting news made  headlines in newspapers around the world and were reported for the general public, such as in the front cover of the New York Times \cite{NYT96}.  A few years later, a similar experiment  was performed at the FERMILAB \cite{Bla97} with 57 events identified, consistent with our perturbative calculations performed before the experiment  \cite{BeBa98} (see also \cite{BD01}). The properties of anti-atoms  are now investigated using ion traps with the purpose to study fundamental symmetries \cite{Eug10,Gro10}. Larger antimatter atoms such as anti-deuterium, anti-tritium, and anti-helium can also be produced in UPCs \cite{Aga11}. Ref. \cite{BeEl10} has made further predictions for the UPC production of muonic, pionic, and other exotic atoms by the coherent photon exchange between nuclei at the Large Hadron Collider (LHC).  These predictions have not yet been observed.

It is also worth noticing that the process of electron-positron production with the capture of the electron in an atomic orbit of one of the nuclei was proposed as a source of beam loss in relativistic colliders already in 1988 in Ref. \cite{Bau88}.  Our estimates at the time was a degradation of a large-Z ion beam within 2 hours at the LHC. This process was further studied  \cite{Klein14} and recent experiments at the LHC \cite{Schau20} have confirmed our expectations \cite{Bau88}. Another process of interest is the production of ortho- and para-positronium in UPCs. Ref. \cite{BN02} has provided a theoretical formalism based on quantum field theory to calculate positronium and other bound-states such as mesons (bound $q\bar q$)  in $\gamma\gamma$- and $\gamma\gamma\gamma$-fusion in UPCs \cite{BBGN16,BGMN17}. 

\section{Light-by-light scattering}
\begin{figure}[h]
\begin{minipage}{10pc}
\includegraphics[width=10pc]{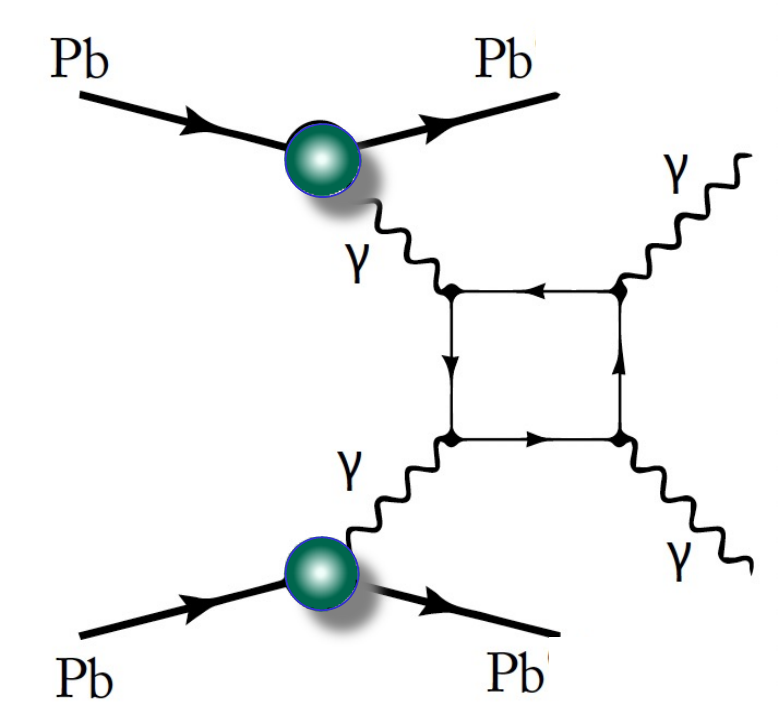}
\caption{\label{lbl}Feynman diagram for light-by-light scattering in UPCs.}
\end{minipage}\hspace{1pc}%
\begin{minipage}{14pc}
\includegraphics[width=14pc]{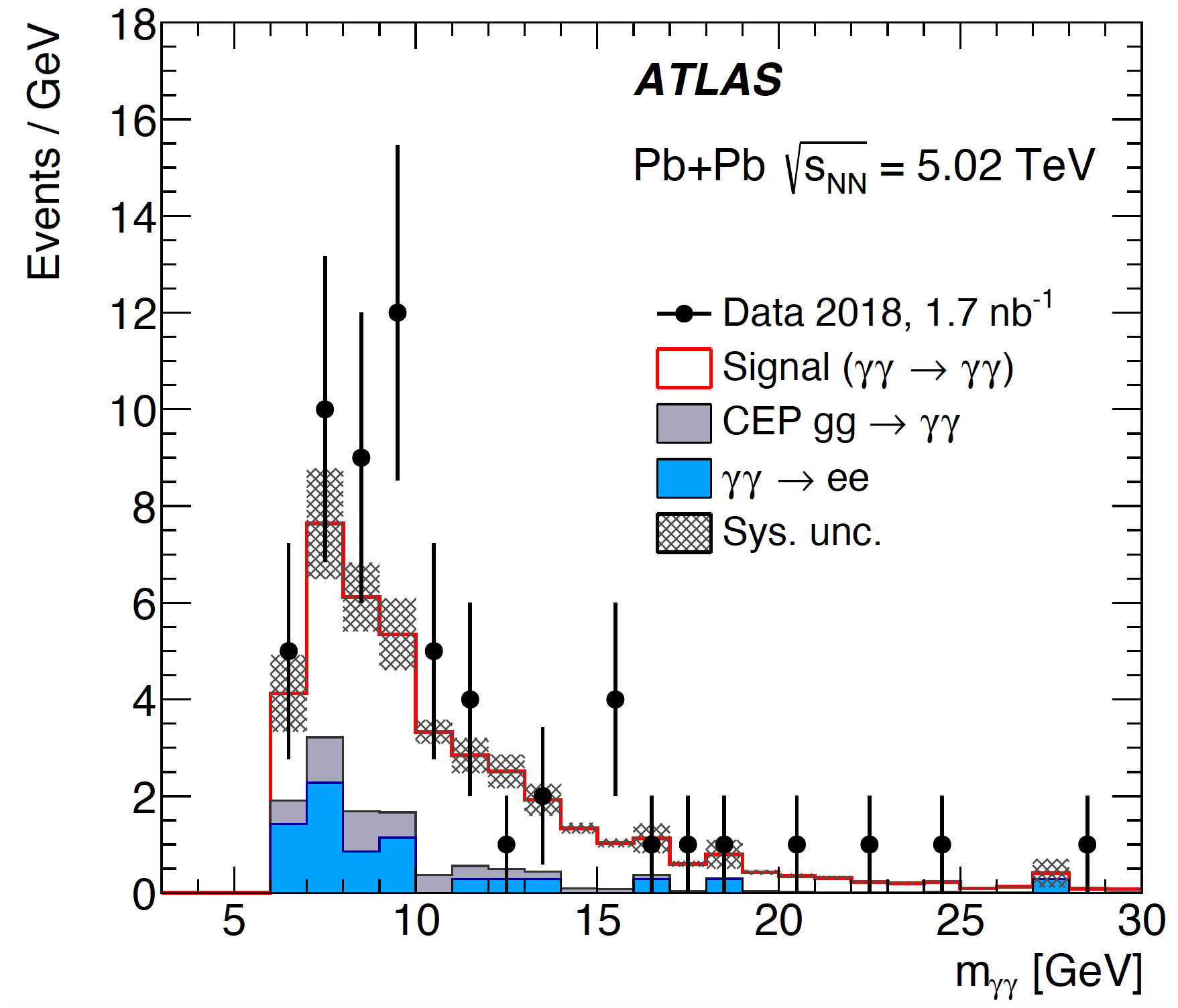}
\caption{\label{lblexp}Light-by-light scattering events observed in UPCs at the ATLAS detector at the LHC \cite{AT17}.}
\end{minipage}\hspace{1pc}%
\begin{minipage}{12pc}
\includegraphics[width=12pc]{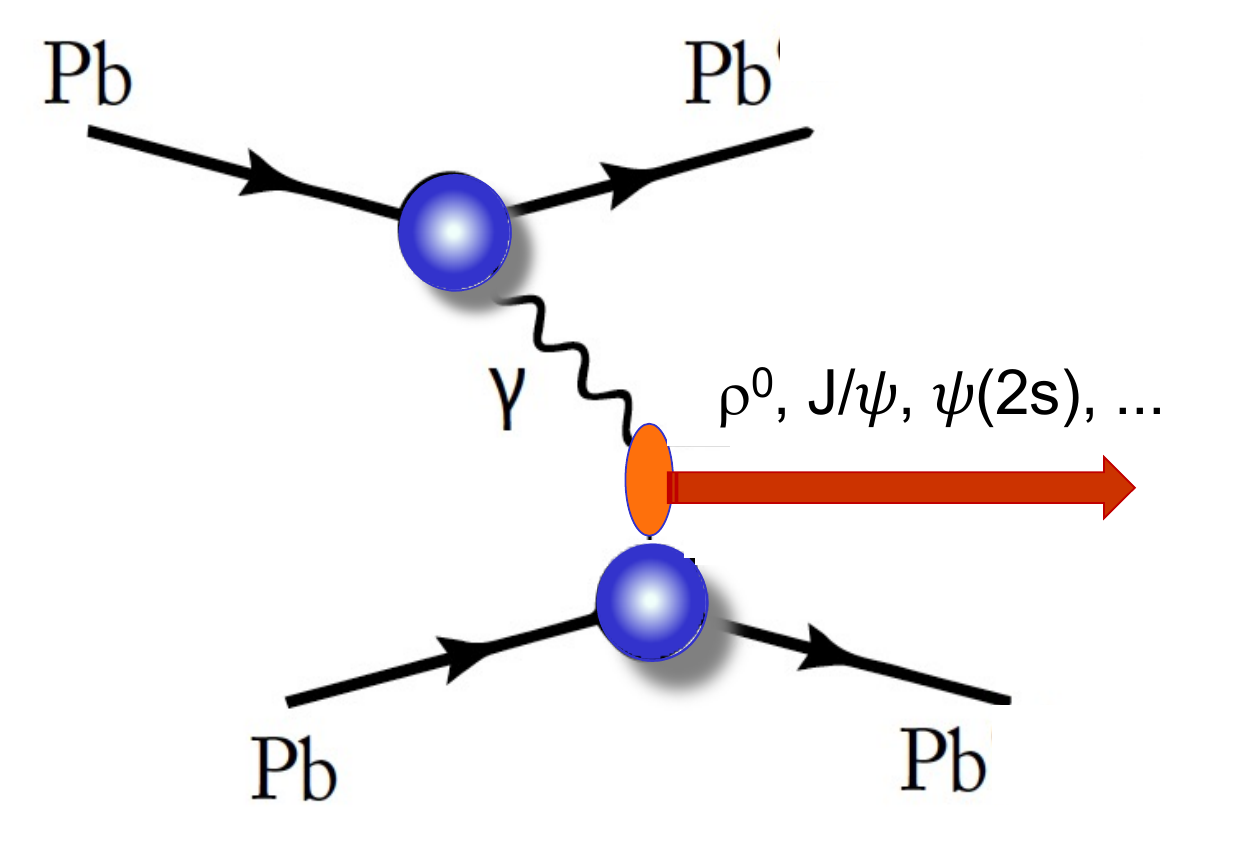}
\caption{\label{jpsi}Schematic diagram for the production of vector mesons in UPCs.}
\end{minipage} 
\end{figure}

The elastic scattering of light by light, $\gamma + \gamma \rightarrow \gamma + \gamma $, can only proceed via the fluctuation of a photon into an $e^+e^-$ pair (Fig. \ref{lbl}). It has a rather small probability and has never been possible to study directly. In Ref. \cite{BeBaZP88} we made the first proposal to use UPC to look for $Z_1+Z_2 \rightarrow Z_1+Z_2 + \gamma+\gamma$. The process would involve two virtual photons scattering via a box-diagram yielding two real photons as a byproduct. We have shown that the calculations based on the Delbr\"uck scattering formalism was very uncertain theoretically \cite{BeBaZP88}. This process was recently observed (2017) in an experiment at the LHC by the ATLAS collaboration \cite{AT17} (Fig. \ref{lblexp}). Such an observation opens the door for the search of physics beyond the Standard Model (SM) because a measured cross section larger than that predicted by the SM model \cite{Klu19} could be a signal of new particles such as axions \cite{GS20,GMR21}.
 
\section{Meson production in UPC} In Ref. \cite{BeBaZP88} the EPN method was generalized to calculate  the production of a bound-particle $X$ in UPCs (Fig. \ref{jpsi}).  One can write the cross section for photon-photon fusion as \cite{BeB88},
\begin{equation}
\sigma_{X}=\int {d\omega_{1}\over \omega_1}{d\omega_{2}\over \omega_2} n_{\gamma}\left(  \omega_{1}\right)  n_{\gamma}\left(
\omega_{2}\right)  \sigma^{X}_{\gamma\gamma}\left(  \omega_{1}\omega_{2}\right)  \;,
\label{eq:two-photon}%
\end{equation} 
where $n_{\gamma}(\omega)$ are the EPNs for a photon with energy $\omega$ and $\sigma
^{X}_{\gamma\gamma}\left(  \omega_{1}\omega_{2}\right)  $ is the photon-photon cross section to produce of particle $X$. 
The photon-photon cross section can be calculated using Low's expression \cite{Low60} based on the detailed balance theorem,
\begin{equation}
\sigma^{X}_{\gamma\gamma}\left(  \omega_{1}\omega_{2}\right)  =8\pi^{2}(2J+1){\frac
{\Gamma_{m_{X}\rightarrow\gamma\gamma}}{m_{X}}}\ \delta\left(  \omega_{1}\omega_{2}
-m^{2}_{X}\right)  \label{Low}%
\end{equation}
where $J$, $m_{X}$, and $\Gamma_{m_{X}\rightarrow\gamma\gamma}$ are the respective spin,
mass and photon-photon $\gamma\gamma$ decay width of particle $X$. The delta-function enforces energy conservation. In Ref. \cite{Bert09} the relevance of different meson models and of exotic  states is discussed thoroughly, including states that have not been considered before.

\subsection{The Higgs} As an interlude, I'd like to mention that in 1989 a proposal emerged to search of the Higgs particle using UPCs at the LHC \cite{Pap89}. The production process is explained by means of Eqs. (\ref{eq:two-photon},\ref{Low}) with appropriate assumptions for the properties of the Higgs. We have made initial estimates in 1988 and obtained a cross section of 1 nb \cite{Bau88}. This is about the same value as the cross section for Higgs production at the LHC through hadronic processes with the advantage that the production of other particles is minimized. However, it was later noticed that another process, the direct photon-photon production of $b\bar b$ pairs, has a much larger cross section. Since the main mechanism of Higgs decay is through  $b\bar b$ pairs, one concludes that the photon-photon production of the Higgs would be swamped in a background of directly produced pairs.  The long sought Higgs particle was finally observed at the LHC in hadronic interactions \cite{BN02,BBGN16,BGMN17}.

\subsection{Exotic mesons}  Multiquark states  such as multiquark molecules of the type $(q\overline{q})(q\overline{q})$, hybrid mesons $(q\overline{q}g)$ and glueballs $(gg)$ are of large interest in meson spectroscopy \cite{Yao06}. UPCs  can contribute to the search of multiquark  resonances through  anomalous $\gamma\gamma$ couplings and  multiquark energy spectrum. UPCs might help to test the predictions of such ``abnormal" states \cite{BN02,BBGN16,BGMN17}.  The $\gamma\gamma$ width is a measure of the charge of  the constituent quarks. Therefore,  the magnitude of the $\gamma\gamma$ coupling is useful to distinguish quark resonances from gluon-dominated resonances (``glueballs''). The absence of meson production via $\gamma\gamma$ fusion is also a useful signal for glueball search \cite{BN02,BBGN16,BGMN17}. In UPCs a glueball can only be formed through annihilation of a $q\overline{q}$ pair into gluon pairs, whereas normal $q\overline{q}$ mesons are created directly.

\subsection{Vector mesons and PDFs.} The production of vector mesons such as $J/\psi$ and  $\Upsilon(1s)$ can also be studied with Eq. (\ref{epn2}) using the appropriate EPN. In 2001 \cite{Goncalves:2001vs} we were the first to propose  this process to constrain  generalized partonic distribution in nuclei, $F_A(x,Q^2)$, for a given momentum fraction $x$. For the real photon induced mechanism we used
\begin{equation}
\left. {d\sigma^{\gamma A \rightarrow VA}\over dt }\right |_{t=0} = {16 \pi \alpha_s^2 (Q^2) \Gamma_{ee}\over 3 \alpha M_V^5} \left[ xF_A(x,Q^2)\right]^2,
\end{equation}
where $\alpha_s(Q^2)$ is the strong interaction coupling evaluated at the perturbative Quantum Chromodynamics (pQCD) factorization scale $Q^2 = W_{\gamma g} ^2$, $M_V$ is the vector meson mass, $\Gamma_{ee}$ is its leptonic decay width and $x=M_V^2/W^2_{\gamma p}$ is   the fraction of the nucleon momentum carried by the gluons.
\begin{figure}[h]
\begin{minipage}{17pc}
\includegraphics[width=17pc]{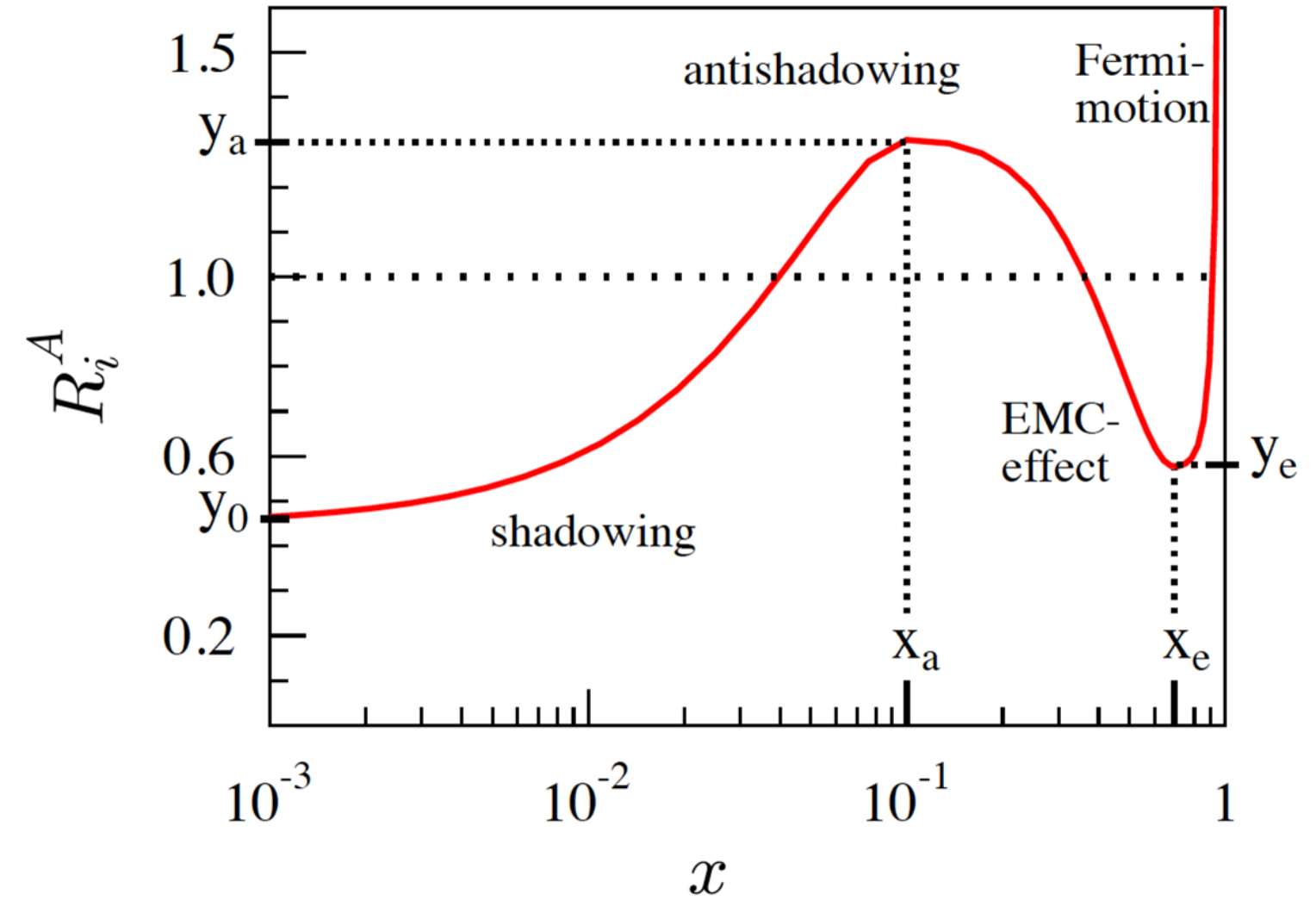}
\caption{\label{emc}Medium modification function displaying numerous effects as a function of the momentum fraction $x$.}
\end{minipage}\hspace{1pc}%
\begin{minipage}{19pc}
\includegraphics[width=19pc]{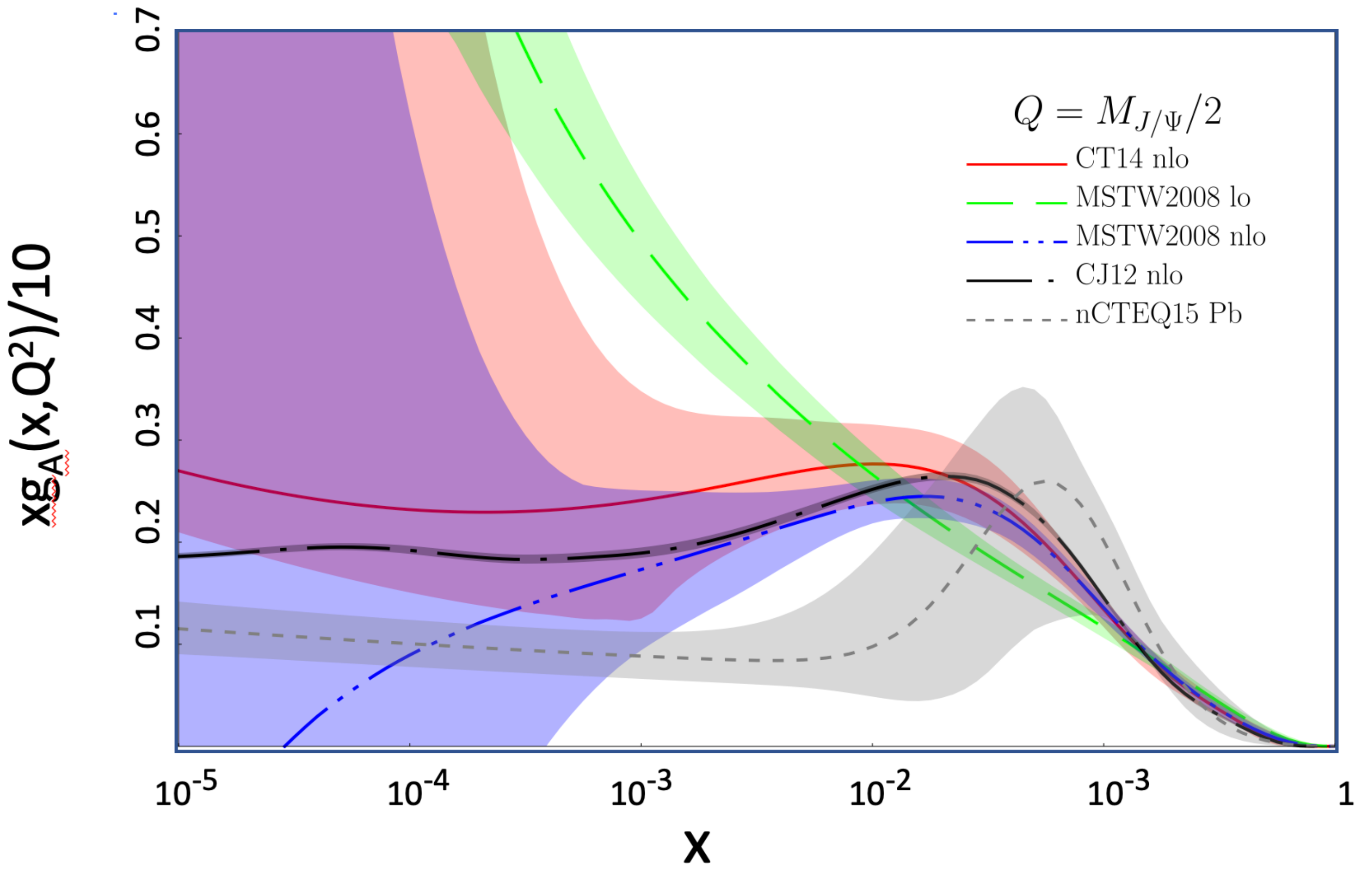}
\caption{\label{pdf}Uncertainties in theoretical compilations of gluon distribution functions (Courtesy of Fred Olness \cite{Ol23}).}
\end{minipage}%

\end{figure}
The nuclear Parton Distribution Function (PDF), $F_{a}^A({\bf r},x,Q^2)$, can be written   as a convolution of a medium modification function $R_{a}^A({\bf r},x,Q^2)$ with the nucleon
PDF, $f_{a}(x,Q^2)$, with the subscript $a$ denoting a parton species and the superscript  $A$ denoting a particular nucleus \cite{Adeluyi:2011rt,Adeluyi:2012ph}, and  ${\bf r}$ is the nucleon position within the nucleus.
Nuclear modifications are included in $R_{a}^A(x,Q^2)$. For $x < 0.04$, one observes  experimentally a shadowing effect, where the nuclear PDFs are smaller compared to the free nucleon distributions, $R_a^A < 1$ (Fig .\ref{emc}). 
In the region $0.04 < x < 0.3$ one observes an anti-shadowing effect,  where $R_a^A > 1$. The EMC effect occurs in the range $0.3 < x < 0.8$. Another enhancement occurs for $x > 0.8$, due to the Fermi motion of nucleons.  The physical principles governing these effects are quite different.
\begin{figure}[h]
\begin{minipage}{18pc}
\includegraphics[width=18pc]{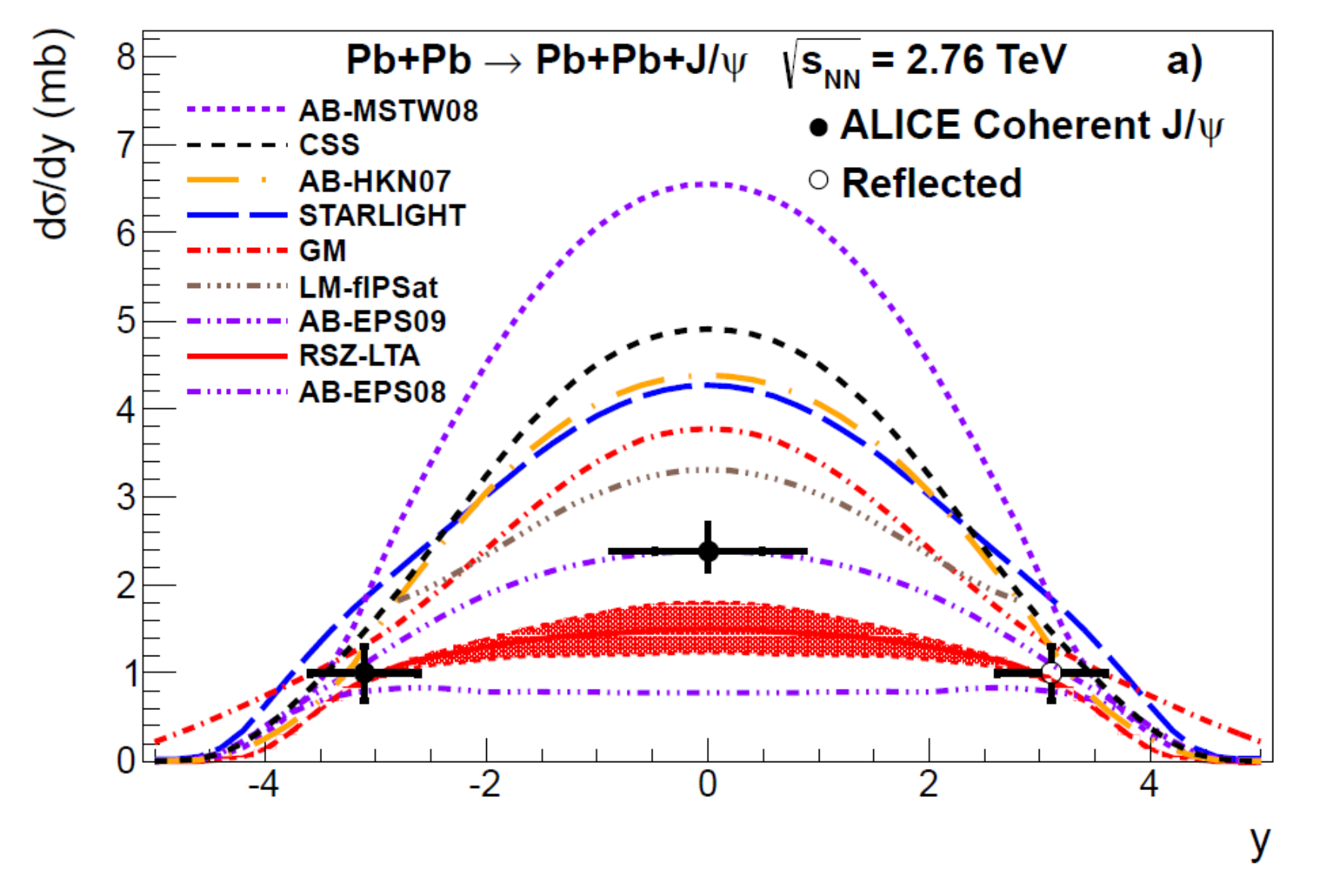}
\caption{\label{pdf1}UPC production of J/$\Psi$ at the LHC as a function of the rapidity $y$, compared to various PDFs (from Ref. \cite{Gru14}).}
\end{minipage}\hspace{2pc}%
\begin{minipage}{18pc}
\includegraphics[width=18pc]{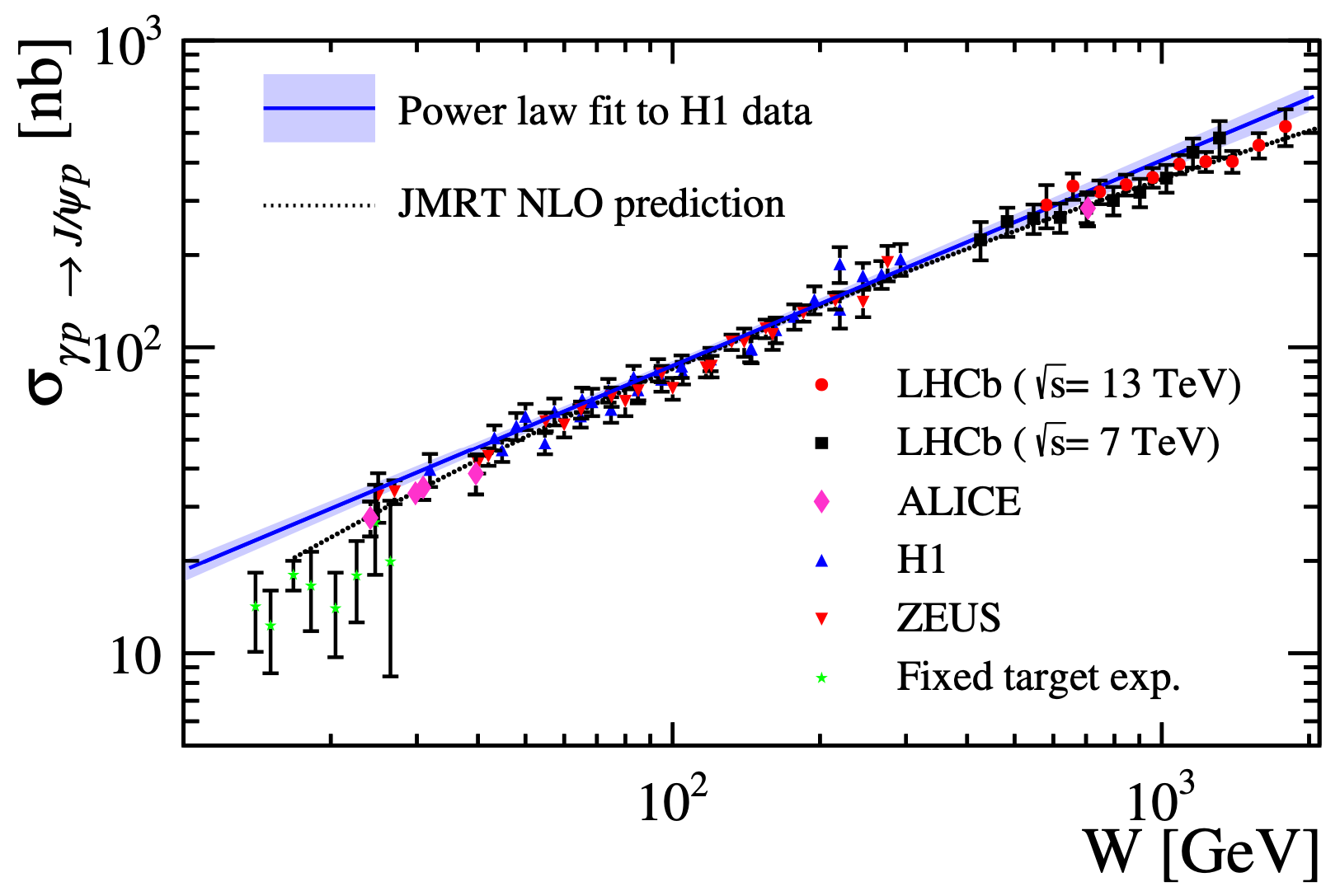}
\caption{\label{pdf2}Compilation of cross sections for UPC production of J/$\Psi$ in different nuclear facilities (from Ref. \cite{Aa18}).}
\end{minipage}\hspace{1pc}%
\end{figure} 
    
In Refs.  \cite{Adeluyi:2011rt,Adeluyi:2012ph} we have studied the impact of different  gluon distributions in $J/\psi$ and  $\Upsilon(1s)$) production in UPCs. It is worthwhile noticing that UPCs with  pPb and PbPb collisions lead to different  production mechanisms  called {\it direct and resolved}. The direct production   implies that the photon interacts directly with the nucleus, while the resolved mechanism implies that the photon fluctuates into a quark-antiquark pair  followed by a hadronic interaction with the nucleus. At leading order the direct production depends on the gluon distributions which are largely uncertain within the nucleus, specially at small $x$ (Fig. \ref{pdf} \cite{Ol23}). The resolved mechanism involves the distributions of light quarks and gluons in both photon and nucleus. 
The quadratic dependence of vector meson production in UPCs increases the sensitivity on gluon distributions of both cross sections and rapidity distributions   \cite{Goncalves:2001vs,Adeluyi:2011rt,Adeluyi:2012ph}.

Our calculations for $J/\Psi$ production with  gluon distributions that incorporate  nuclear gluon shadowing  \cite{Adeluyi:2011rt,Adeluyi:2012ph} is in good agreement with the ALICE experimental data \cite{Abe13,Abe13b,Gru14}, as shown  in Fig. \ref{pdf1}.  It is now clear that $J/\Psi$ and $\Upsilon$  photoproduction  at the LHC are a powerful tool to study nuclear gluon shadowing in the  region  $x<10^{-3}$.  A compilation of cross sections at different facilities is shown in Fig. \ref{pdf2}  showing a beautiful power law (see Ref. \cite{Aa18} for more details).

\section{Physics beyond the Standard Model}
In this brief review I have shown how genuinely simple ideas developed in the 1980s and 1990s have become a fruitful research area at the relativistic  hadron colliders. Unexpected phenomena induced in UPCs have been verified experimentally such as double giant resonances, the dawn of studies with anti-atoms, light-by-light scattering, beam losses due to bound-electrons, probing parton distribution functions,  and search for physics beyond the standard model. Could $\gamma\gamma \rightarrow {\rm graviton}$ \cite{ANP20}, multi-quark systems \cite{Esp21}, or axion-like particles \cite{Ad18,GS20,GMR21} be found in this way? The jury is out on whether this will be possible with further advances in new accelerators, beam physics and new detection techniques.     

\bigskip

{\bf Acknowledgments}

 This work was partially supported by the U.S. DOE grant DE-FG02-08ER41533 and NSF grant no. 2114669 (Accelnet IANN-QCD https://www.iann-qcd.org).

\medskip

{\bf References}




\begin{thebibliography}{9}

\bibitem{BeBa85} Bertulani C A and Baur G 1985 Relativistic Coulomb collisions  {\it Nucl. Phys.} A 442 p  739.

\bibitem{Fer24} Fermi E 1924 On the Theory of Collisions between Atoms and Electrically Charged Particles {\it Zeit. Phys.} 29 p 315.

\bibitem{Fer25} Fermi E 1925 Sulla Teoria Dell' Urto Tra Atomi E Corpuscoli Elettici {\it Nuovo Cimento} 2 p 143.

\bibitem{Wei34} Weizs\" acker C F Radiation due to collisions of very fast electrons {\it Z. Phys. 88} p 612.

\bibitem{Wil34} Williams E J 1934 Nature of the High Energy Particles of Penetrating Radiation and Status of Ionization and Radiation Formulae {\it Phys. Rev.} 45 p 724; Williams E J 1935 {\it K. Dan. Vidensk. Selsk. Mat.-Fys. Medd.} p 13 no. 4.

\bibitem {Sch49} Schwinger J S  1949 Quantum electrodynamics. III. The electrodynamic properties of the electron {\it Phys. Rev.} 76 p 790.

\bibitem {Bau08}Baur G 2008 Exchange of high and low energy photons in ultraperipheral relativistic heavy-ion collisions {\it Nucl. Phys. B}  179 p 129.

\bibitem{BeB88} Bertulani C A and Baur G 1988 Electromagnetic processes in relativistic heavy ion collisions {\it Phys. Reports} 163 p 299.

\bibitem {KGS97}Krauss F, Greiner M,  Soff G 1997 Photon and gluon induced processes in relativistic heavy ion collisions {\it Prog. Part. Nucl. Phys.} 39 p 503.

\bibitem{BHT02} Baur G, Hencken K, Trautmann D,  Sadovsky S and Kharlov Y  2002 Coherent $\gamma\gamma$ and $\gamma A$ interactions in very peripheral collisions at relativistic ion colliders {\it Phys. Reports} 364 p 359.

\bibitem{BKN05} Bertulani C A,  Klein S R and Nystrand J 2005  Physics of ultra-peripheral nuclear collisions {\it Annu. Rev. Nucl. Part. Sci.} 55 p 271.

\bibitem {BHT07}Baur G, Hencken K and Trautmann D 2007 Electron-positron pair production in ultrarelativistic heavy ion collisions {\it Phys. Reports} 453 p 1.

\bibitem {Ba08}Baltz A J et al. 2008 The physics of ultraperipheral collisions at the LHC  {\it Phys. Reports} 458 p 1.

\bibitem {Ba08b}Baltz A J 2008 Evidence for higher order QED in e+ e-pair production at RHIC {\it Phys. Rev. Lett.} 100  062302.

\bibitem {Nyst08}Nystrand J for the ALICE Collaboration 2008 Photon-Induced Physics with Heavy-Ion Beams in ALICE {\it Nucl. Phys. B (Proc.
Suppl.)} 179 p 156.

\bibitem{BeB86} Baur G and Bertulani C A 1986 Electromagnetic processes in relativistic heavy ion collisions {\it Nucl. Phys. A} 458 p 725. 

\bibitem{ABS95} Aumann T,  Bertulani C A and Suemmerer K 1995 Neutron removal in peripheral heavy ion collisions {\it Phys. Rev. C} 51 p 416.

\bibitem{ASBK96} Aumann T, Suemmerer K, Bertulani C A and Kratz J V 1996 Excitation of the two-phonon giant dipole resonance in 238U studied via inclusive measurements of neutron-removal cross sections {\it Nucl. Phys. A} 599 p 321.

\bibitem{Aum97} Aumann T, Borbtzky K, Stroth J, Wajda E, Blaich Th, Cub J, and Holeczek Jacek 1997 Coulomb fragmentation and Coulomb fission of relativistic heavy-ions and related nuclear structure aspects {\it Acta Physica Polonica. B} 1 p  375.

\bibitem{BKL20} Bertulani C A, Kucuk Y and Lozeva R 2020 Fission of relativistic nuclei with fragment excitation and reorientation {\it Phys. Rev. Lett.} 124 132301.

\bibitem{BCW98} Baltz A J, Chasman C and  White S N 1998 Correlated forward–backward dissociation and neutron spectra as a luminosity monitor in heavy-ion colliders {\it Nucl. Inst. Meth. Phys. Res. A} 41 p 1.

\bibitem{BeB94} Bertulani C A and Baur G 1994 (March) Relativistic heavy ion physics without nuclear contact  {\it Physics Today} p. 22. 

\bibitem{BeBa86} Baur G and Bertulani C A 1986 Multiple electromagnetic excitations of giant dipole phonons in relativistic heavy ion collisions {\it  Phys. Lett. B} 174 p 23. 

\bibitem{BeBau86} Baur G and Bertulani C A 1986 Multistep fragmentation of heavy ions in peripheral collisions at relativistic energies {\it Phys. Rev. C} 34  p 1654. 

\bibitem{Sch93}  Schmidt R et al. 1993 Electromagnetic excitation of the double giant dipole resonance in 136Xe {\it Phys. Rev. Lett.} 70 p 1767.  

\bibitem{Rit93} Ritman J et al 1993 First observation of the Coulomb-excited double giant dipole resonance in  208 Pb  via double-gamma decay {\it Phys. Rev. Lett.} 70 p 533.

\bibitem{Em94} Emling 1994 Electromagnetic excitation of the two-phonon giant dipole resonance {\it Prog. Part. Nucl. Phys.} 33 p 729.

\bibitem{ABE98} Aumann T, Bortignon P F and  Emling H 1998 Multiphonon giant resonances in nuclei {\it Ann. Rev. Nucl. Part. Sci.} 48 p 351.

\bibitem{BP99} Bertulani C A and Ponomarev V 1999 Microscopic studies of the double giant resonance  {\it Phys. Reports} 321 p 139.

\bibitem{BeB89} Baur G and Bertulani C A 1989 Electromagnetic processes in relativistic heavy ion collisions {\it Nucl. Phys. A} 505 p 835.

\bibitem {Bau88} Baur G and Bertulani C A, 1988 Electromagnetic physics at relativistic heavy ion colliders: for better and for worse {\it Z. Phys. A}  330 p 77.

\bibitem{Ad18} Adam W et al - CMS collaboration 2018 Evidence for light-by-light scattering and searches for axion-like particles in ultraperipheral PbPb collisions at $\sqrt{s}_{NN}=5.02$ TeV {\it Phys. Lett. B} 797 p 303.

\bibitem{BeBa87} Baur G and Bertulani C A 1987 Electromagnetic production of heavy leptons in relativistic heavy ion collisions {\it Phys. Rev. C} 35 p 836.

\bibitem {MB94} Munger C and Brodsky S 1934 Production of relativistic antihydrogen atoms by pair production with positron capture {\it Physical Review D} 49  p 3228.

\bibitem{Bau96} Baur G et al 1996 Production of Antihydrogen {\it Physics Letters B} 368 p 3.

\bibitem{NYT96} Browne M W  1996 Physicists Manage to Create The First Antimatter Atoms {\it The New York Times} January 5. 

\bibitem{Bla97} Blanford G et al 1998 Observation of Atomic Antihydrogen {\it Phys. Rev. Lett.} 80 p 3037. 

\bibitem{BeBa98}  Bertulani C A and Baur G  1998 Theoretical calculation of antihydrogen production and accuracy of the equivalent photon approximation {\it Phys. Rev. D} 58 034005.

\bibitem{BD01} Bertulani C A and Dolci D 2001 Pair production with capture in peripheral collisions with heavy ions {\it  Nucl. Phys. A} 683 p 635.

\bibitem{Klein14} Klein S R 2014 Heavy ion beam loss mechanisms at an electron-ion collider {\it Phys. Rev.: Accelerators and Beams} 17 121003.

\bibitem{Schau20} Schaumann, Jowett J M, Bahamonde Castro C, Bruce R, Lechner A and  Mertens T 2020 Bound-free pair production from nuclear collisions and the steady-state quench limit of the main dipole magnets of the CERN Large Hadron Collider {\it Phys. Rev.: Accelerators and Beams} 23 121003

\bibitem{Eug10} Reich E S 2010 Antimatter held for questioning {\it Nature News} 11 17.

\bibitem{Gro10} Grossman L 2010 The Coolest Antiprotons {\it Phys. Rev. Focus} 26 p 1.

\bibitem{BN02} Bertulani C A and Navarra F 2002 Two- and three-photon fusion in relativistic heavy ion collisions {\it Nucl. Phys. A} 703  p 861. 

\bibitem{BBGN16} Moreira B D, Bertulani C A,  Gon\c calves V P and F. S. Navarra 2016 Production of exotic charmonium in gamma-gamma interactions at hadronic colliders {\it Phys. Rev. D} 94 094024.

\bibitem{BGMN17} Bertulani C A, Gon\c calves V P, Moreira B D and  Navarra F S 2017 Production of exotic charmonium in ultra-peripheral heavy ion collisions {\it Eur. Phys. J.}  137 06019. 

\bibitem{Aga11} Agakishiev H et al 2011 Observation of the antimatter helium-4 nucleus {\it Nature} 473 p 353.

\bibitem{BeEl10} Bertulani C A and Ellermann M  2010 Production of exotic atoms at the CERN Large Hadron Collider {\it Phys. Rev. C} 81 044910.

\bibitem{BeBaZP88} Baur G and Bertulani C A 1988 $\gamma\gamma$ physics with peripheral relativistic heavy ion collisions {\it Z. Phys. A} 330 p 77.

\bibitem{AT17} Atlas collaboration 2017 Evidence for light-by-light scattering in heavy-ion collisions with the ATLAS detector at the LHC {\it Nature Physics} 13 p 852.

\bibitem{Klu19} Kłusek-Gawenda M2019  Importance of mesons in light-by-light scattering in ultraperipheral lead-lead collisions at the LHC {\it EPJ Web Conf.} 199 05004.

\bibitem{GS20} Gon\c calves and Sauter W K 2020 Exclusive axionlike particle production by gluon - induced interactions in hadronic collisions, {\it Phys. Lett. B} 811 135981.

\bibitem{GMR21} Gon\c calves V P, Martins D E  and Rangel M S 2021 Searching for axionlike particles with low masses in pPb and PbPb collisions {\it Europ. Phys. J C}  81 p: 522. 

\bibitem{Low60}Low F E  1960 Proposal for Measuring the $\pi^0$ Lifetime by $\pi^0$ Production in Electron-Electron or Electron-Positron Collisions {\it Phys. Rev.} 120 p 582.

\bibitem{Bert09} Bertulani C A 2009 Probing two-photon decay widths of mesons at energies available at the CERN Large Hadron Collider (LHC) {\it Phys. Rev. C} 79 047901.

\bibitem {Pap89}Papageorgiu E 1989 Coherent Higgs Production in Relativistic Heavy Ion Collisions {\it Phys. Rev. D} 40 p 92.

\bibitem{Yao06} Yao W-M et al 2006 Particle Data Group: Non-quark antiquark Mesons {\it J. Phys. G} 33 p 1.

 \bibitem{Goncalves:2001vs} Gon\c calves V P and Bertulani C A 2002 Peripheral heavy ion collisions as a probe of the nuclear gluon distribution {\it Phys.\ Rev.\  C} 65  054905. https://arxiv.org/abs/hep-ph/0110370
  
  \bibitem{Adeluyi:2011rt}  Adeluyi A and Bertulani C A 2011 Gluon distributions in nuclei probed at the CERN Large Hadron Collider {\it Phys.\ Rev.\ C} 84  024916.

\bibitem{Adeluyi:2012ph}  Adeluyi A and Bertulani C A 2012 Constraining Gluon Shadowing Using Photoproduction in Ultraperipheral pA and AA Collisions {\it Phys.\ Rev.\ C} 85 044904.

\bibitem{Ol23} Olness F 2023 Private Communication.
  
 \bibitem{Abe13} Abelev B et al 2013 (ALICE Collaboration) Coherent  photoproduction in ultra-peripheral Pb-Pb collisions at $\sqrt{s_{NN}}= 2.76$ TeV {\it Phys. Lett. B} 718 p 1273.

\bibitem{Abe13b} Abelev B et al 2013 (ALICE Collaboration) Charmonium and e$^+$e$^-$ pair photoproduction at mid-rapidity in ultra-peripheral Pb-Pb collisions at $\sqrt{s_{NN}}=2.76$ TeV {\it Eur. Phys. J. C }73 p 2617.
  
\bibitem{Gru14} De Gruttola D (ALICE Collaboration) 2014 $J/\Psi$ photoproduction in Pb-Pb and p-Pb ultra-peripheral collisions with ALICE at LHC  {\it Nucl. Phys. A} 926 p 136.

\bibitem{Aa18} Alan et al - The LHCb collaboration 2018 Central exclusive production of J/$\Psi$ and $\Psi$(2S) mesons in pp collisions at $\sqrt{s}=13$ TeV {\it J. High Energy Phys.} 10 p 167. 

\bibitem{Esp21} Esposito A,  Manzari C A, Pilloni A and Polosa A D 2021 Hunting for tetraquarks in ultraperipheral heavy ion collisions {\it Phys. Rev. D} 104 114029.

  
\bibitem{ANP20} Ahern S, Norbury J W and Poyser W J 2000 Graviton production in relativistic heavy-ion collisions {\it Phys. Rev. D} 62 116001.



\end{thebibliography}
\end{document}